\documentclass[journal]{IEEEtran}

\usepackage{cite}      % Written by Donald Arseneau
\usepackage[shortlabels]{enumitem}

\usepackage{graphicx}
\usepackage{amsmath}
\usepackage{amsfonts}
\usepackage{url}
\usepackage{color}
\usepackage[normalem]{ulem}
\usepackage{mathabx,graphicx}
\usepackage{epstopdf}
\usepackage{hyperref}
\usepackage{amsfonts}

\newcommand{\be}{\begin{equation}}
\newcommand{\ee}{\end{equation}}
\newcommand{\bea}{\begin{eqnarray}}
\newcommand{\eea}{\end{eqnarray}}
\newcommand{\df}{{\operatorname d}}

\begin{document}

\title{Effects of geomagnetic field perturbations on the power supply of transoceanic fiber optic cables}

\author{Antonio Mecozzi,~\IEEEmembership{Fellow,~OSA, Fellow,~IEEE}
	%\thanks{Manuscript received \today. The author acknowledge financial support from the Italian Government through project Innovating City Planning
		%through Information and Communication Technologies (INCIPICT), and the PRIN 2017 project Fiber Infrastructure for Research on Space-Division
		%Multipled Transmission (FIRST).}% <-this % stops a space
	\thanks{A. Mecozzi is with the Department of Physical and Chemical Sciences,
		University of L'Aquila, L'Aquila 67100, Italy}}

%\markboth{Journal of Lightwave Technology}{Shell
	%\MakeLowercase{\textit{et al.}}: Bare Demo of IEEEtran.cls for
	%Journals}

%\specialpapernotice{Invited Tutorial}
% \IEEEspecialpapernotice{(Invited Paper)}

\maketitle

\begin{abstract}
There is a growing concern that a big coronal mass ejection event will induce perturbations on the power supply of fiber optic transoceanic cables that may produce a global internet blackout. In this paper we give the expression of the voltage variations that a transient change of the geomagnetic field induces on the voltage of the power supply of a transoceanic fiber optic cable. We show that the transient voltage change is proportional to the magnitude of the magnetic field deviations and not to its time derivative as a direct application of Faraday's law would imply, and this suggests design criteria to protect transoceanic fiber optic systems against big geomagnetic storm events. The presented analysis also enables the classification of existing systems into some that are less sensitive to the weakening of the geomagnetic field occurring during strong geomagnetic storms and others that are more prone to experience an outage when a weakening of the geomagnetic field occurs.
\end{abstract}

\begin{IEEEkeywords}
Optical communications, Transoceanic optical cables
\end{IEEEkeywords}

\section{Introduction}

The interaction of the geomagnetic field with charged particles during coronal mass ejection (CME) may induce significant electromagnetic perturbations, known as geomagnetic storms. There is a growing concern that a big CME event will induce perturbations on fiber optic transoceanic systems that may end up in a global internet blackout \cite{Jyothi:21}. 

In this paper, we present a theory of the interaction between the power line feeding a transoceanic fiber optic system and a time-varying magnetic field. We show that the combined effect of the frequency dependent attenuation of the electromagnetic perturbation in the conducting seawater and the reflection from the bottom, has the effect that the induced electric field in not proportional to the derivative of the magnetic field as a direct application of Faraday's law would suggest, but to the amplitude of the magnetic field perturbation. The theory predicts that the effective area that was used to quantify the effect of the magnetic perturbation on the induced voltage variations becomes linearly proportional to the duration of the perturbation, and this result is used to check the validity of the analysis with experimental data extracted from the literature \cite{Medford:89}. The predicted proportionality of the voltage variations to the amplitude of the magnetic field perturbations has also very practical implications. In the presence of a sudden drop of the magnetic field, like the ones that are likely to occur during large magnetic storms \cite{Mohanty:16,Cliver:13}, the induced voltage variations will follow the magnetic field perturbation, and undergo an increase or a decrease depending on the direction of the electric current. %This result suggests how to reduce, with a specific design, the sensitivity of fiber optic plans to possible future strong geomagnetic storm events, and provides a simple criterion to estimate the sensitivity of an existing system to strong perturbations of the geomagnetic field.
This result, beside providing a simple criterion to estimate the vulnerability of an existing fiber optic plan, provides design criteria to reduce the sensitivity of newly deployed systems to possible future strong geomagnetic storm events.

\section{The model} 

In a transoceanic optical cable, the in-line equipment is fed by a positive and a negative (or zero) DC voltage from the cable terminals using a wire that is grounded at its ends. Effectively, the return line is replaced by the earth. The voltage balance is
\be V_1 + V_+ - V_\mathrm{sys} - V_ - = V_2, \ee
where $V_1$ is the potential of the earth at the transmitter (or the receiver) side,  $V_+$ is the positive voltage of the feed at the transmitter (or the receiver)  side, $V_\mathrm{sys}$ is the voltage drop due to the line resistance and the voltage drop of the in-line devices, $V_-$ is the absolute value of the negative voltage of the feed at the receiver (or the transmitter) side, and $V_2$ is the potential of the earth at the receiver (or the transmitter) side. Writing
\be V_+-V_- %= V_\mathrm{sys} +(V_2-V_1) 
= V_\mathrm{sys} +\Delta V, \ee
where $\Delta V = V_2-V_1$, one sees right away that the required feeding voltage is equal to the voltage drop of the system plus the difference of the earth potential at the receiver and the transmitter side (or vice versa). In a static situation, the earth voltage is uniform. This is not the case when the geomagnetic field that surrounds the earth changes with time. Under the action of a time varying magnetic fields, electric fields are generated by Faraday's law. For slowly varying fields like the ones involved in geomagnetic storm events, the quasi static approximation is always valid, so that an electric potential can still be defined. In this case, however, the presence of a time-varying magnetic field produces a non-uniform earth potential. Therefore, if one assumes that the voltage drop of the system is constant, the voltage variations of the power supply of transoceanic transmission systems monitor the variations of the difference of the earth potentials at the transmitter and receiver sides. 

In order to analyze this problem quantitatively, let us write Faraday, Amp\`ere-Maxwell and Ohm laws in the ocean, in the quasi static approximation, as follows
\bea \vec \nabla \times \vec E &=& - \frac{\partial \vec B}{\partial t}, \label{180} \\
 \vec \nabla \times \vec B &=& \mu_0 \vec J  + \mu_0 \epsilon \frac{\partial \vec E}{\partial t}, \label{190}  \\
 \vec J &=& \sigma \left(\vec E + \vec v \times \vec B \right). \label{200} \eea
Here, $\sigma$ is the electrical conductivity of the water, and we neglect the very small perturbation introduced by the cable wire. Let us consider the case $v = 0$ first, and a time varying magnetic field, and assume the quasi static approximation neglecting the Maxwell term proportional to the time derivative of the electric field in the Amp\`ere-Maxwell law (\ref{190}). Using then Eq. (\ref{200}) into Eq. (\ref{190}) we obtain
 \be \vec \nabla \times \vec B = \mu_0 \sigma \vec E. \ee
 Inserting the traveling wave solution with $\omega >0$ and $\vec B(-\omega) = \vec B^*(\omega)$ and $\vec E(-\omega) = \vec E^*(\omega)$
\bea \vec B &=& \vec B_0 \exp\left(i \vec k \cdot \vec r - i \omega t \right), \\
\vec E &=& \vec E_0 \exp\left(i \vec k \cdot \vec r - i \omega t \right), \eea
we obtain
\bea \vec k \times \vec E_0 &=& \omega \vec B_0, \\
\vec k \times \vec B_0 &=& - i \mu_0 \sigma \vec E_0. \eea
The vectors $\vec E_0$, $\vec B_0$ and $\vec k$ are a right-handed orthogonal set of vectors. We have
\be \vec k \times \left(\vec k \times \vec E_0\right) = - i \mu_0 \sigma \omega \vec E_0, \ee
that is 
\be - k^2 \vec E_0 = - i \mu_0 \sigma \omega \vec E_0 \Longrightarrow k^2 = i \mu_0 \sigma \omega. \ee

\begin{figure}[ ht]
\centering\includegraphics[width=9cm]{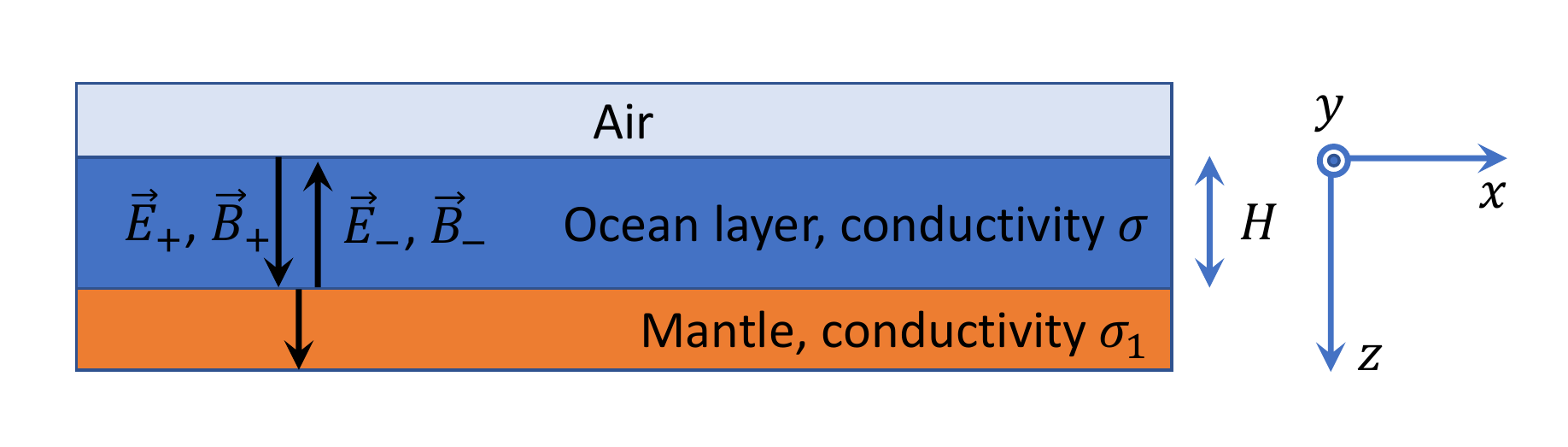}
\caption{Schematic of the model.} \label{Fig0}
\end{figure}

Let us now approximate the ocean as a flat water layer of width $H$ and resistivity $\sigma$, with air on top and lying on a flat semi-infinite substrate (the mantle) of resistivity $\sigma_1$ \cite{Wait:62}, and use a reference frame defined by the base unit vectors $\hat e_z$ directed downwards, $\hat e_x$ parallel to $\vec E_0$, and $\hat e_y$ parallel to $\vec B_0$ to complete a right-handed reference frame, see Fig. \ref{Fig0}. Assume that the cable is lying at the interface of water and the flat semi-infinite substrate, at depth $z = H$. Assume that the electromagnetic field is propagating in the water along the $z$ direction so that $\vec k = \pm k \hat e_z$. The above equation yields two attenuated traveling wave solutions proportional to
\be \exp\left(i k z - i \omega t \right) = \exp(\pm \kappa \, z - i \omega t).\ee
where
\be \kappa = \sqrt{- i \mu_0 \sigma \omega}, \ee
and we use the convention that the real part of $\kappa$ is positive. With this choice if we take the minus sign we have the attenuated progressive  wave propagating in the direction $\hat e_z$
\be \vec E_+ = \vec E_0  \exp(- \kappa \, z - i \omega t)= \vec E_0 \exp\left[(i-1) \frac{z}{\delta(\omega) } - i \omega t \right] \ee
and if we take the positive sign the attenuated regressive wave propagating in the direction opposite to $\hat e_z$
\be \vec E_- = \vec E_0 \exp(\kappa \, z - i \omega t)= \vec E_0 \exp\left[-(i-1) \frac{z}{\delta(\omega) } - i \omega t \right], \ee
where
\be \delta(\omega) = \sqrt{\frac{2}{\mu_0 \sigma \omega}} \ee
is the skin attenuation depth. Let us define the Fourier amplitudes as $E(z,t) = E(z) \exp(-i \omega t)$ and $B(z,t) = B(z) \exp(-i \omega t)$, assume
\bea \vec E(z) &=& E(z) \hat e_x, \label{Ez0} \\
\vec B(z) &=&  B(z) \hat e_y,\label{Bz0}  \eea
and decompose that the fields as a combination of forward and backward propagating components, we have
\bea E(z) &=& E_+ \exp(-\kappa \, z) + E_- \exp(\kappa \, z ), \label{Ez} \\
B(z) &=& B_+ \exp(-\kappa \, z) - B_- \exp(\kappa \, z). \label{Bz} \eea
Using Eqs. (\ref{180}) and (\ref{190}) in the quasi static approximation we obtain
\be \frac{E_+}{B_+} = \frac{E_-}{B_-} = \frac{i k}{\mu_0 \sigma} = \sqrt{\frac{-k^2}{\mu_0^2 \sigma^2}} = \sqrt{\frac{-i \omega}{\mu_0 \sigma}}  = \frac {Z} {\mu_0}, \ee
where $Z= E_+/H_+ = \sqrt{-i \omega \mu_0/\sigma}$ is the impedance of the ocean water, so that Eq. (\ref{Bz}) becomes
\be B(z) =  \sqrt{\frac{\mu_0 \sigma}{-i \omega}} \left[ E_+ \exp(-\kappa \, z) - E_- \exp(\kappa \, z)\right], \label{Bz1} \ee
%
%
%\be \frac{E(z)}{B(z)} = \sqrt{\frac{-i \omega}{\mu_0 \sigma}} \, \frac{E_+ \exp(-\kappa \, z) + E_- \exp(\kappa \, z )}{E_+ \exp(-\kappa \, z) - E_- \exp(\kappa \, z)}. \ee
%
 Defining the reflection coefficient $\Gamma$ of an incident wave on the substrate as
\be \Gamma = \frac{E_- \exp(\kappa \, H)}{E_+ \exp(-\kappa \, H)}, \label{Gamma0} \ee
and using the preservation of the parallel components of $\vec E$ and $\vec B$ at the interface and the absence of a backward propagating field returning from the semi-infinite medium, it is possible to show that $\Gamma$ has the expression
\be \Gamma = \frac{Z_1 -Z}{Z_1+Z} = \frac{(\sigma_1)^{-1/2} - (\sigma)^{-1/2}}{(\sigma_1)^{-1/2} + (\sigma)^{-1/2}}, \label{Gamma}
\ee
where we used that the impedance of the mantle is $Z_1=  \sqrt{-i \omega \mu_0/\sigma_1}$. The use in Eq. (\ref{Ez}) and in Eq. (\ref{Bz1}) calculated for $z = 0$ of the expression for $E_-$ obtained from Eq. (\ref{Gamma0}), that is $E_- = \Gamma \exp(-2 \kappa \, H) E_+$, yields
\bea  \frac{E(z)}{E_+} &=&   \exp(-\kappa \, z) + \Gamma \exp[\kappa \, (z-2H)], \\
\frac{B(0)}{E_+} &=&  \sqrt{\frac{\mu_0 \sigma}{-i \omega}} \left[ 1 - \Gamma \exp(-2 \kappa \, H)\right]. \eea
The ratio of the first to the second of the equations above gives
\be \frac{E(z)}{B(0)} = \sqrt{\frac{-i \omega}{\mu_0 \sigma}} \, \frac{\exp[\kappa \,(H- z)]+\Gamma \exp[-\kappa \, (H-z)]}{\exp(\kappa \, H) - \Gamma \exp(-\kappa \, H)},  \ee
or, using Eq. (\ref{Gamma}) and after straightforward algebra\footnote{This equation generalizes the combination of Eqs. (27) and (28) of ref. \cite{Wait:62} that gives $Z = \mu_0 E(0)/B(0)$, and becomes identical to that for $z = 0$, after the replacements $\sigma \to \sigma_1$, $\sigma_1 \to \sigma_3$ and $H \to h_1$, and with $-i$ replaced by $i$ because of the different Fourier transform definition.}
\be \frac{E(z)}{B(0)} \! = \! \sqrt{\frac{-i \omega}{\mu_0 \sigma}}  
\frac{\cos[\kappa \, (H-z)]+\left(\sigma_1/\sigma \right)^{1/2} \sinh[\kappa \, (H-z)]}{\sinh(\kappa \, H)+\left(\sigma_1/\sigma \right)^{1/2}  \cosh(\kappa \, H)}. \label{ZZ} \ee
For $z = H$, Eq. (\ref{ZZ}) becomes
\be E(H) = \sqrt{\frac{-i \omega}{\mu_0 \sigma}} \, \frac{B(0)}{\sinh(\kappa \, H)+\left(\sigma_1/\sigma \right)^{1/2}  \cosh(\kappa \, H)}, \ee
where the magnetic field $B(0)$ is the total magnetic field evaluated at the interface between water and air.  Using now Eq. (\ref{Ez0}) for $z = H$ and Eq. (\ref{Bz0}) for $z = 0$, and that $\hat e_x = \hat e_y \times \hat e_z$, we obtain the final result \cite{Boteler:89}
\be \vec E(H) = \sqrt{\frac{-i \omega}{\mu_0 \sigma}} \, \frac{\vec B(0) \times \hat e_z}{\sinh(\kappa \, H)+\left(\sigma_1/\sigma \right)^{1/2}  \cosh(\kappa \, H)}. \ee
Outside the water, the propagation constant is $k = \omega/c$ so that the wavelength is $\lambda = c T$ where $T = 2 \pi/\omega$ is the period of the perturbation. For $T$ larger than one second, the wavelength is larger than $300,000$ km, larger than the size of the magnetosphere. This means that the plane wave approximation cannot be applied in air and the quantity of interest is the total magnetic field at the air-water interface, not a forward-propagating magnetic field in air, which cannot be defined outside the framework of a plane wave approach. For the generation of a planar wave in the ocean layer, instead, we need to assume that the amplitude and phase of the geomagnetic field is uniform on the water surface over the length scale set by the wavelength in the water $2 \pi/\mathrm{imag}(k) = 2 \pi \delta(\omega)$. For time scales $\Delta \tau = 2 \pi/\omega$ ranging from $10^3$ to $10^5$ s, which are typical time scales of geomagnetic field fluctuations, $\delta(\omega)$ lies between $9$ to $90$ kilometers for $\sigma = 3 \, \Omega^{-1}$m$^{-1}$.

Let us now consider a cable parametrized by the curvilinear coordinate $s$ and be $\df \vec s$ its increment. We will allow that the direction of the magnetic field varies along the cable over a length scale long enough that the plane wave approximation still holds in the water layer, and we also allow a variability of $H$ over the same length scale. Let us assume that the return path of the circuit goes deep into the earth mantle in a region where $\vec E \simeq 0$. The induced electromotive force in the circuit acts only in the cable section and its component at frequency $\omega$ is
\be {\cal E}  =  \sqrt{\frac{-i \omega}{\mu_0 \sigma}} \int_0^L  \frac{\hat e_z \times  \df \vec s  \cdot \vec B(0)}{\sinh(\kappa \, H)+\left(\sigma_1/\sigma \right)^{1/2}  \cosh(\kappa \, H)}, \label{E2} \ee
where we used that $\vec B(0) \times \hat e_z \cdot \df \vec s = \hat e_z \times  \df \vec s  \cdot \vec B(0)$. Let us now use a local reference frame defined by the base unit vectors $\hat e_z$, $\hat e_x$ and $\hat e_y$ where $\hat e_x$ is defined by $\df \vec s = \hat e_x \df s$ and $\hat e_y$ by the property that it completes a right-handed orthogonal reference frame together with the vertical $\hat e_z$ pointing downwards, and assume that the cable is lying at depth $z = H(s)$. In this reference frame Eq. (\ref{E2}) becomes
\be {\cal E} =  \sqrt{\frac{-i \omega}{\mu_0 \sigma}} \int_0^L  \frac{ B_y(\omega) \df s}{\sinh(\kappa \, H)+\left(\sigma_1/\sigma \right)^{1/2}  \cosh(\kappa \, H)}, \ee
where $B_y(\omega) = \hat e_y  \cdot \vec B(0)$, and we made explicit the dependence of the magnetic field on the angular frequency $\omega$ and implicit the fact that the magnetic field is referred to the interface ocean-air, that is at $z = 0$.  A constant intensity in the cable requires that the total voltage $V$ needed for the nominal current flow in the system be constant in the presence of the perturbation $ {\cal E}$, that is $V + \Delta V + {\cal E} = V$, so that the voltage change $\Delta V$ in response to the generation of the electromotive force $\cal E$ is $\Delta V = - {\cal E}$, that is
\be \Delta V = - \sqrt{\frac{-i \omega}{\mu_0 \sigma}} \int_0^L  \frac{B_y(\omega) \df s}{\sinh(\kappa H) +\left(\sigma_1/\sigma \right)^{1/2}  \cosh(\kappa \, H)}. \label{DeltaVomega2} \ee
When the current flows in the circuit in the positive direction of $s$ and of $\hat e_x$, $V$ is positive and hence a positive $\Delta V$ corresponds to an increase of the voltage $V$ of the power supply. On the other hand, when the current flows in the opposite direction, $V$ is negative and hence a positive $\Delta V$ corresponds to a decrease of the absolute value of $V$.

\section{Approximate analytical expressions}

%For the time scale involved, the exponent at the denominator can be replaced by one so that
%
%\be \Delta V = - \int_0^L  \sqrt{\frac{-i \omega}{\mu_0 \sigma}} \, \frac{1+\Gamma}{1 - \Gamma +(1+\Gamma) \kappa \, H(s)} \, B_y(\omega) \df s. \ee
%
%For shorter time scales instead, 
Equation (\ref{DeltaVomega2}) shows that $\Delta V(\omega) = 0$ for $\omega = 0$ and hence $\Delta V(t) = 0$ for a constant $B_y(t)$. We are interested into events which have a finite time duration over a constant background. We may therefore separate the constant background using
\be B_y(t) = B_{y,0} + \Delta B_y(t), \ee
where $\Delta B_y(t) \to 0$ for $|t| \to \infty$ sufficiently fast that its spectrum is finite.  Because $\Delta V(\omega)$ does not change if $B_y(t)$ is replaced by $\Delta B_y(t)$ by subtracting a constant bias, it is legitimate to replace in Eq. (\ref{DeltaVomega2}) $B_y$ by $\Delta B_y$ in the following analysis.

In order to obtain some useful analytical relation between the voltage and the magnetic field variations, let us expand the denominator of Eq. (\ref{DeltaVomega2}) the arguments of the hyperbolic functions as
\be \Delta V(\omega) = - \sqrt{\frac{-i \omega}{\mu_0 \sigma}} \int_0^L  \frac{\Delta B_y(\omega) \df s}{\alpha + \alpha^3/6 + \left(\sigma_1/\sigma \right)^{1/2} (1 + \alpha^2/2)}, \ee
where $\alpha = \kappa \, H$. If we assume that $\Delta B_y(\omega)$ is bandwidth limited to the frequency $\Omega = 2 \pi/\Delta \tau$, then we have $|\alpha| < 1$ for $ \Delta \tau > 2 \pi \mu_0 \sigma H^2$. Assuming $\sigma = 3 \, \Omega^{-1}$ m$^{-1}$ for the value of the conductivity of the ocean water, and $H = 4$ km, we obtain $\Delta \tau > 380$ s. For events that fulfill this condition we may then write
\be \Delta V(\omega) = - \sqrt{\frac{-i \omega}{\mu_0 \sigma}} \int_0^L  \frac{\Delta B_y(\omega) \df s}{\alpha  + \left(\sigma_1/\sigma \right)^{1/2}}. \label{DeltaVomega1} \ee
The value of the conductivity of the substrate is much smaller than that of the water, of the order of $0.01 \, \Omega^{-1}$ m$^{-1}$ or less. Assuming the value $\sigma_1 = 0.01 \, \Omega^{-1}$ m$^{-1}$ the second term at the denominator is dominant for $\mu_0 \sigma (2 \pi/\Delta \tau) H^2 > (\sigma_1/\sigma) $, that is for $\Delta \tau < (\sigma/\sigma_1) \mu_0 \sigma 2 \pi H^2$ which, with our values of the parameters is $\Delta \tau < 10^5$ s. For $\Delta \tau > 380$ s and $\Delta \tau < 10^5$ both the first order expansion and the neglecting of 1 are legitimate and we have
\be \Delta V(\omega) = - \int_0^L  \frac{ \Delta B_y(\omega) \df s }{\mu_0 \sigma H(s)}, \ee
and in time domain
\be \Delta V(t) = -  \frac{L}{\mu_0 \sigma} \left[\frac 1 L \int_0^L  \frac{ \Delta B_y(t) \, \df s}{H(s)}\right] .  \ee
For constant $H(s) = H$ and constant $\Delta B_y(t)$ we then obtain the simple relation
\be \Delta V(t) = -  \frac{L }{\mu_0 \sigma H} \, \Delta B_y(t), \label{DeltaVt} \ee
Assuming $H = 4$ km and $\sigma = 3 \, \Omega^{-1}$m$^{-1}$, we obtain $\Delta V/(L \, \Delta B_y) \simeq 66.3$ mV/(Mm nT) (Mm stands for megameter, a handy unit for transoceanic system lengths).  Notice that, contrary to what could be anticipated from a direct application of the Faraday's law, the voltage variations are proportional to the amplitude of the magnetic field perturbations, not to its time derivative. This can be explained by the fact that the derivative becomes multiplication by $-i \omega$ in frequency domain, and that the coupling between the induced electric field and the derivative of the magnetic field is inversely proportional to $\sqrt{-i \omega \mu_0 \sigma}$ and the enhancement originated by the reflection of the substrate is in turn, for high enough frequencies, inversely proportional to $\sqrt{-i \omega \mu_0 \sigma} \, H$, so that the effective coupling between the voltage and  the magnetic field perturbations turns out to be independent of $\omega$.

If we use in Eq. (\ref{DeltaVt}) the dimensionless time $\tau = t/\Delta \tau$, we obtain
\be \Delta V(\tau) = -  \frac{L \, \Delta B_y(\tau)}{\mu_0 \sigma H} , \label{DeltaVtau} \ee
which shows that a consequence of the dependence of the voltage variations on the amplitude of the magnetic field perturbations is that the proportionality coefficient is independent on the time scale $\Delta \tau$ of the magnetic field perturbation.

For very low frequencies, corresponding to $\Delta \tau > 10^5$ s, we are in the opposite limit in which we may instead neglect the term proportional to $H$ in Eq. (\ref{DeltaVomega1}). If we do so, we obtain
\be \Delta V(\omega) = - \sqrt{\frac{-i \omega}{\mu_0 \sigma_1}} \int_0^L  \Delta B_y(\omega) \df s. \ee
An elegant and very useful formal expression in time domain of the above equation can be obtained in terms of the derivative of order one-half of the magnetic field (the derivative of order one-half is a linear operator that applied twice produces a first derivative \cite{Podlubny:17}) 
\be \Delta V (t) = - \sqrt{\frac{1}{\mu_0 \sigma_1}}  \int_0^L   \left(\frac{\df}{\df t} \right)^{1/2} \, \Delta B_y(t) \df s, \label{DeltaVt1} \ee
In this case, the use of the dimensionless time $\tau = t/\Delta \tau$ gives
\be \Delta V (\tau)= - \sqrt{\frac{1}{\mu_0 \sigma_1 \Delta \tau}} \int_0^L  \left(\frac{\df}{\df \tau} \right)^{1/2} \, \Delta B_y(\tau) \df s.  \label{DeltaVtau1} \ee
For a gaussian $\Delta B_y(\tau)$ the ratio between the maxima of the half-derivative and the derivative is about 3, and it is in all cases of the order of unit. Equation (\ref{DeltaVtau1}) shows an inverse proportionality of $\Delta V (\tau)$ with the square root of the time scale of the perturbation $\Delta \tau$.

Direct application of Faraday's law suggests the definition of the effective area $A$ \cite{Medford:81,Medford:89}
\be \Delta V(t) = - A \frac{\df \Delta B_y(t)}{\df t}, \ee
that is, using the dimensionless time $\tau = t/\Delta \tau$, 
\be A = -  \Delta \tau \Delta V(\tau) \left[ \frac{\df \Delta B_y(t)}{\df t} \right]^{-1}. \label{A} \ee
Using for $V(\tau)$ the expression given by Eq. (\ref{DeltaVtau}), valid for $\Delta \tau> 380$ s and $\Delta \tau < 10^5$ s, and expanding $A$ to lowest order in $\Delta \tau$, we obtain that $A$ is linearly dependent on the time scale of the magnetic field perturbation $\Delta \tau$
\be A %-\left[ \Delta \tau \int_0^L  \frac{ \Delta B_y(\tau) \, \df s}{\mu_0 \sigma H(s)}\right]  \left[\left(\frac{\df}{\df \tau} \right) \, \Delta B_y(\tau)\right]^{-1} 
= A_0 \Delta \tau, \ee
with proportionality coefficient
\bea A_0 &=& \int_0^L  \frac{\df s}{\mu_0 \sigma H(s)} \left[\left(\frac{\df}{\df \tau} \right) \, \ln \Delta B_y(\tau)\right]^{-1} \nonumber \\
&=&  \frac{L}{\mu_0 \sigma H} \left[\left(\frac{\df}{\df \tau} \right) \, \ln \Delta B_y(\tau)\right]^{-1}, \label{A0} \eea
where the second equality holds for constant $H(s) = H$ and $\Delta B_y(\tau)$. Using in Eq. (\ref{A}) the expression for $V(\tau)$ given by Eq. (\ref{DeltaVtau1}), valid for $\Delta \tau > 10^5$, we obtain that $A$ is proportional to the square root $\sqrt{\Delta \tau}$ of the time scale of the magnetic field perturbation
\be A = A_1 \sqrt{\Delta \tau}, \ee
with proportionality coefficient
\be A_1 = L \sqrt{\frac{1}{\mu_0 \sigma_1}}  \left[ \frac{\left(\df/\df \tau \right)^{1/2} \, \Delta B_y(\tau)} {\left(\df/\df \tau \right) \, \Delta B_y(\tau)} \right]. \ee

\section{Comparison with literature data}

\begin{figure}[ ht]
\centering\includegraphics[width=6cm]{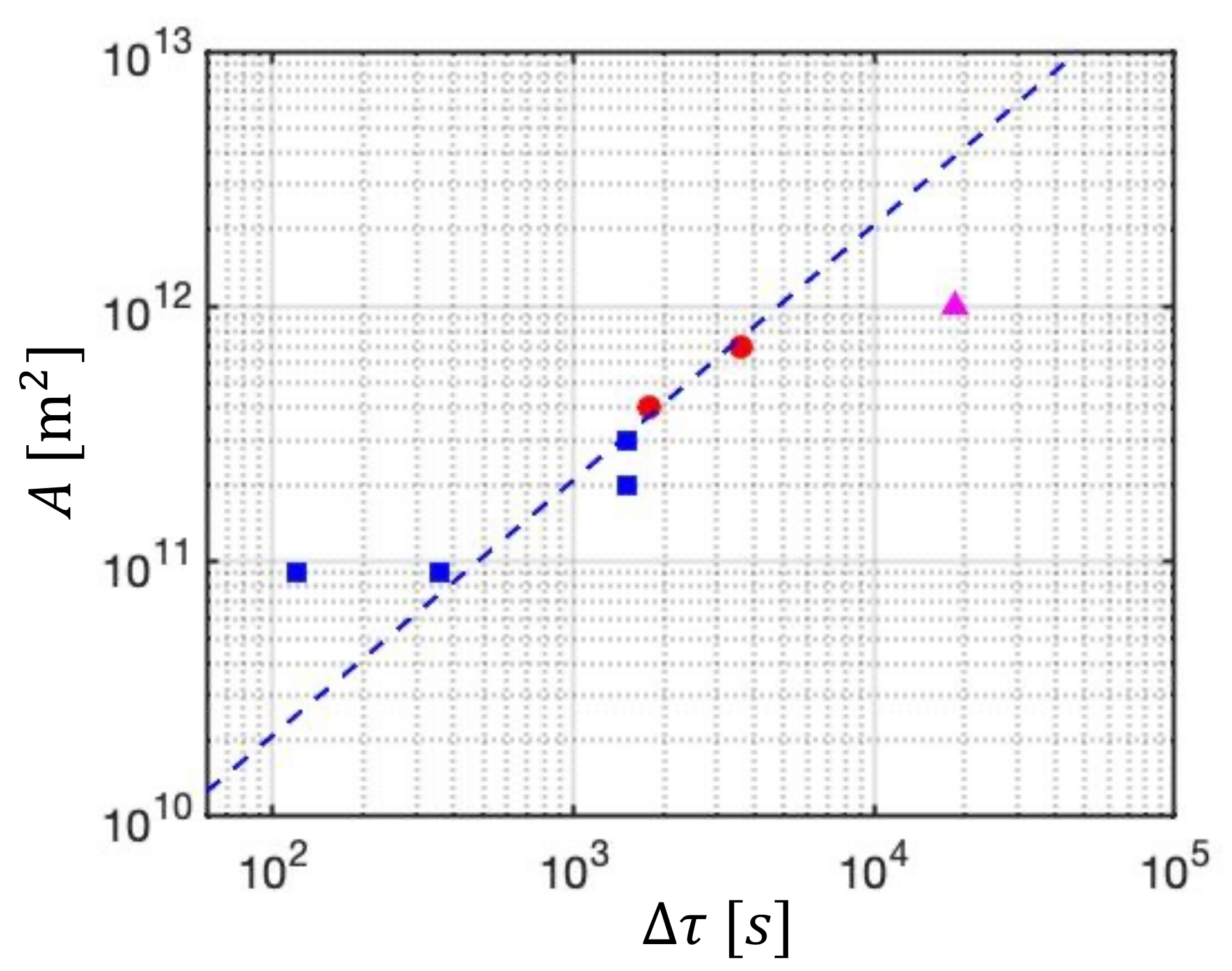}
\caption{Values taken from Table 1 of ref. \cite{Medford:89} of $A$ in square meters vs. the time change of the magnetic field $\Delta \tau$ is seconds for geomagnetic storms events affecting the TAT-8 system (blue squares),  the TAT-7 system (red dots), and the TAT-6 system (magenta triangle). Dashed blue line, plot of the scaling $A = A_0 \,  \Delta \tau$ (see text).} \label{Fig1b}
\end{figure}

In Fig. \ref{Fig1b} we plot the values taken from Table 1 of ref. \cite{Medford:89} of $A$ in square meters vs. the time change of the magnetic field $\Delta \tau$ in seconds for geomagnetic storms events affecting the TAT-8 system (blue squares),  the TAT-7 system (red dots), and the TAT-6 system (magenta triangle).  A rigorous comparison between the $A$ obtained with our theory and that defined in Refs. \cite{Medford:81} and \cite{Medford:89} is of course impossible because of the qualitative definition of $\Delta \tau$ and the presence of the derivative of $\ln(B_y)$ with respect to $\tau=t/\Delta \tau$ in Eq. (\ref{A0}). However, it is reasonable to assume that the absolute value of the derivative of $\ln(B_y)$ is of the order of one (for instance, the derivative of the $\ln \mathrm{sech}(\tau)$ is $-\tanh(\tau)$ which spans the interval from $-1$ to 1), so that an estimate of the scaling of $A$ with the event duration $\Delta \tau$ may be obtained by replacing in Eq. (\ref{A0}) the derivative of the logarithm of $B_y$ with $1$. Using then $\sigma = 6$ $\Omega^{-1}$m$^{-1}$ for the electrical conductivity of the seawater, $L = 6300$ km and $H = 4$ km we obtain $A_0 = 2.1 \, 10^9$ m$^2$s$^{-1}$. The plot of $A = A_0 \Delta \tau$ is reported as a dashed blue line in the figure, showing good agreement with the experimental points, with the exception of two outliers for small and large values of $\Delta \tau$. 

% For the value of an observation on TAT-6 reported in Table 1 of ref. \cite{Medford:89}, $\Delta \tau = 320$ minutes, that is $\Delta \tau = 19200$s, using 1 as the ratio of the half-derivative to the derivative, and $\sigma_1 = 0.15$ $\Omega^{-1}$m$^{-1}$ for the effective electrical conductivity of the ocean mantle, we obtain for $L = 6300$ km of the TAT-6 system the value $A = 2.0 \cdot 10^{-12}$, in good agreement with the observed value (see table 1 of ref. \cite{Medford:89}) of $A = 10^{12}$ m$^2$. The factor two difference may be attributed to the definition of $\Delta \tau$ in ref. \cite{Medford:89}. 

%\begin{figure}[ ht]
%\centering\includegraphics[width=7cm]{Fig1a}
%\caption{Values taken from Table 1 of ref. \cite{Medford:89} of $A$ in square meters vs. the time change of the magnetic field $\Delta \tau$ is seconds for geomagnetic storms events affecting the TAT-8 system (blue squares),  the TAT-7 system (red dots), and the TAT-6 system (magenta triangle). Dashed line, plot of the predicted scaling $A = A_0 \,  \Delta \tau^{1/2}$ with coefficient $A_1 = 10^{12}/19200$ m$^2/$s$^{1/2}$ extracted from the value reported in Table 1 of ref. \cite{Medford:89} for the TAT-6 event.} \label{Fig1a}
%\end{figure}

Let us consider a cable whose path is a straight line east-west, and use a three dimensional right-handed frame with origin on the ocean surface and base unit vectors $\hat e_y$ horizontal pointing north, $\hat e_z$ vertical oriented downwards, and $\hat e_x$ horizontal parallel to the cable and pointing west. In this reference frame, a positive $\Delta B_y(t)$ produces a negative $\Delta V$. In both TAT-6 and TAT-8 systems the current flows west-east, and hence the voltage $V$ is negative and a negative $\Delta V$ corresponds to an increase of the absolute value of the voltage. Equation (\ref{DeltaVt}) predicts for a positive $\Delta B_y(t)$ a negative $\Delta V(t)$, which corresponds in our case to an increase of the voltage $V$, and this is consistent with the experimental observations that show that an increase of the south-north magnetic field component always corresponds to an increase of the absolute value of the voltage \cite{Medford:81,Medford:89}.

Let us now perform a more detailed comparison between the prediction of the theory and the experimental observations reported in Refs. \cite{Medford:81} and \cite{Medford:89}. 

\begin{figure}[ ht]
\centering\includegraphics[width=9cm]{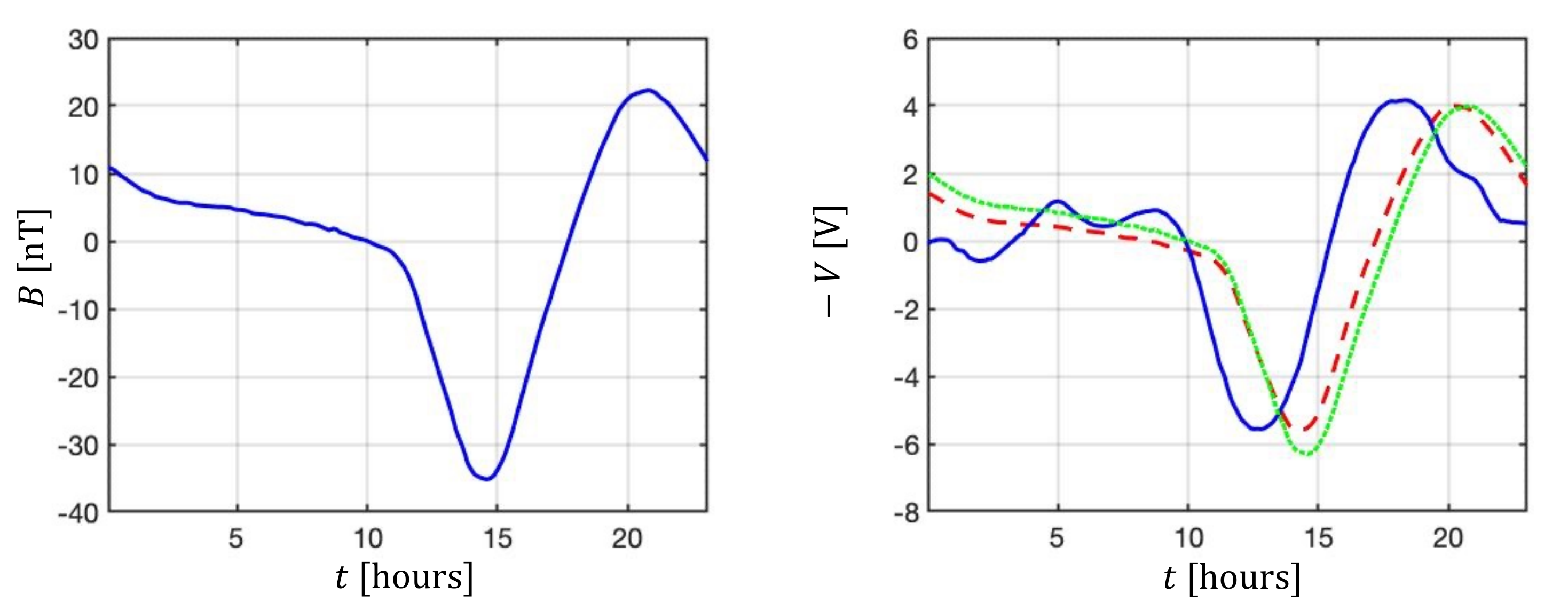}
\caption{Left panel: South-north magnetic field horizontal component averaged over five days, digitized from Fig. 2 of ref.  \cite{Medford:81}. Right panel: the voltage variation in the TAT-6 system obtained from Fig. 3 of the same reference (solid blue line), and the prediction of Eq. (\ref{DeltaVomega2}) with $\sigma = 6 \, \Omega^{-1}$m$^{-1}$ and $\sigma_1 = 0.01 \, \Omega^{-1}$m$^{-1}$ (dashed red line), and of Eq. (\ref{DeltaVt}) with $\sigma = 7 \, \Omega^{-1}$m$^{-1}$ (dotted green line).} \label{Fig7}
\end{figure}

The left panel of Fig. \ref{Fig7} shows the south-north magnetic field horizontal component averaged over five days, digitized from Fig. 2 of ref.  \cite{Medford:81}. The right panel of the same figure shows by a solid blue line line the voltage variation in the TAT-6 system, of length $L = 6300$ km, obtained from Fig. 3 of the same reference, and by a dashed red line the prediction of Eq. (\ref{DeltaVomega2}) with $\sigma = 6 \, \Omega^{-1}$m$^{-1}$ and $\sigma_1 = 0.01 \, \Omega^{-1}$m$^{-1}$, and by a dotted green line the prediction of its approximated version Eq. (\ref{DeltaVt}) with $\sigma = 7 \, \Omega^{-1}$m$^{-1}$. 

\begin{figure}[ ht]
\centering\includegraphics[width=9cm]{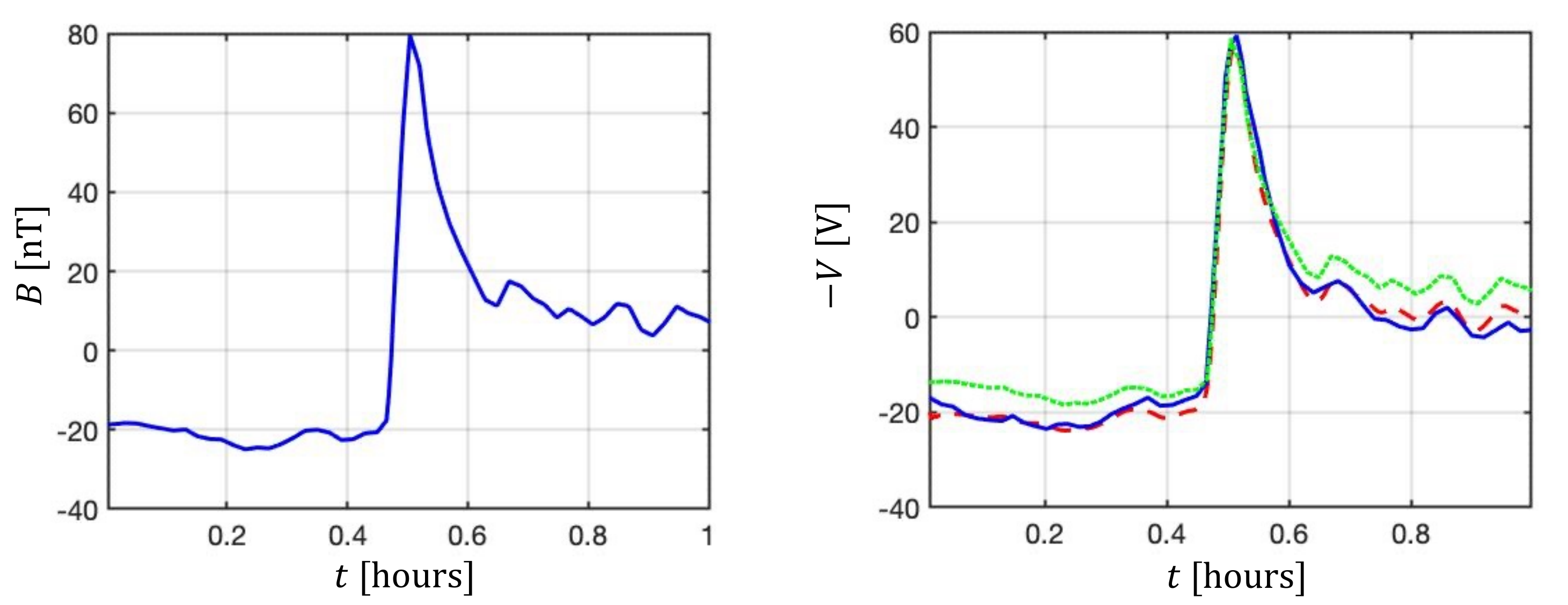}
\caption{Left panel: South-north magnetic field horizontal component digitized from Fig. 2 of ref.  \cite{Medford:89}. Right panel: the voltage variation in the TAT-8 system also obtained from Fig. 2 of ref.  \cite{Medford:89} (solid blue line), and the prediction of Eq. (\ref{DeltaVomega2}) with $\sigma = 1.5 \, \Omega^{-1}$m$^{-1}$ and $\sigma_1 = 0.001 \, \Omega^{-1}$m$^{-1}$ (dashed red line) and of Eq. (\ref{DeltaVt}) with $\sigma = 1.7 \, \Omega^{-1}$m$^{-1}$ (dotted green line)} \label{Fig6}
\end{figure}

Figure \ref{Fig6} compares the results of the theory with the observations recorded during the event reported in Fig. 2 of ref. \cite{Medford:89}. The right panel shows good agreement of the voltage variation observed experimentally (solid blue line) with the predictions of Eq. (\ref{DeltaVomega2}) using $\sigma = 1.5 \, \Omega^{-1}$m$^{-1}$ and $\sigma_1 = 0.001 \, \Omega^{-1}$m$^{-1}$ (dashed red line) and with the predictions of Eq. (\ref{DeltaVt}) using $\sigma = 1.7 \, \Omega^{-1}$m$^{-1}$ (dotted green line). 

\begin{figure}[ ht]
\centering\includegraphics[width=9cm]{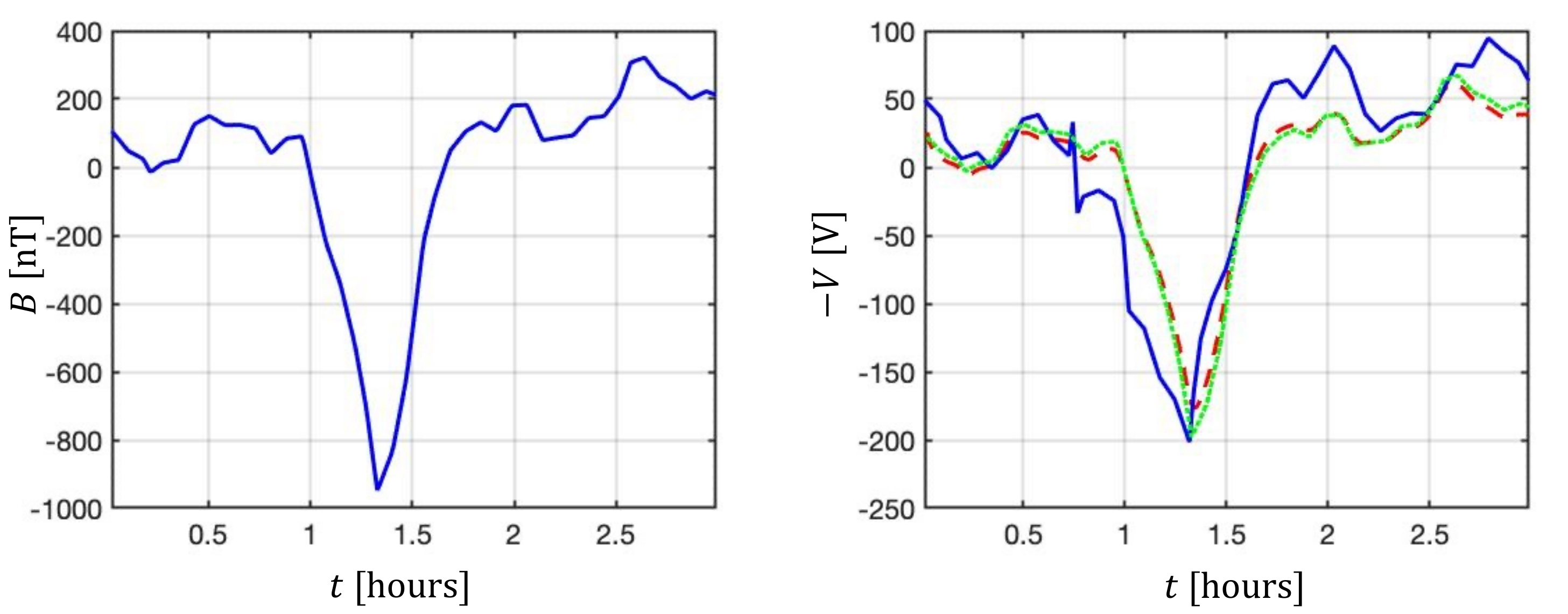}
\caption{Left panel: East-west magnetic field horizontal component digitized from Fig. 4 of ref.  \cite{Medford:89}. Right panel: the voltage variation in the TAT-8 system also obtained from Fig. 4 of ref.  \cite{Medford:89} (solid blue line), and the prediction of Eq. (\ref{DeltaVomega2}) with $\sigma = 6 \, \Omega^{-1}$m$^{-1}$ and $\sigma_1 = 0.01 \, \Omega^{-1}$m$^{-1}$ (dashed red line) and of Eq. (\ref{DeltaVt}) with $\sigma = 6 \, \Omega^{-1}$m$^{-1}$ (dotted green line).} \label{Fig8}
\end{figure}

Figure \ref{Fig8} compares the results of the theory with the event reported in Fig. 4 of ref.  \cite{Medford:89}. In this case, the right panel shows good agreement of the voltage variation observed experimentally (solid blue line) with the predictions of Eq. (\ref{DeltaVomega2}) using $\sigma = 6 \, \Omega^{-1}$m$^{-1}$ and $\sigma_1 = 0.01 \, \Omega^{-1}$m$^{-1}$ (dashed red line), and with the predictions of Eq. (\ref{DeltaVt}) using $\sigma = 6 \, \Omega^{-1}$m$^{-1}$ (dotted green line), using the east-west component of the magnetic field, which shows a magnetic field variation about three times larger than the south-north component. Notice that the geometry of the TAT-8 cable, which runs southwest to northeast, is such that the south-north and east-west magnetic field components have components of the same sign on the direction orthogonal to the cable.

As a general comment, the spread of the value of the conductivity $\sigma$ and $\sigma_1$ in Figs. \ref{Fig7}--\ref{Fig8} is caused by the fact that the magnetic field to be used in Eq. (\ref{DeltaVomega2}) is the spatial average of the component orthogonal to the local cable direction, while we have used a single component, south-north in Figs. \ref{Fig7} and \ref{Fig6} and east-west in Fig. \ref{Fig8}, at a fixed location. The three cases analyzed however show very clearly that the shape of the voltage variation is not proportional to the time derivative of the magnetic field variation, which would produce positive voltage variations in the leading edge and negative in the trailing, but to the magnitude of its deviations. This is evident for the good agreement between the observations and the shape predicted by Eq. (\ref{DeltaVt}), especially for the shortest time scale events of Figs. \ref{Fig6} and \ref{Fig8}, where the approximations over which Eq. (\ref{DeltaVt}) is based are more accurate.

We have shown above that in both TAT-6 and TAT-8 systems the sign of the electromotive force is consistent with the prediction of the theory. It is interesting to notice that the choice of the direction of the current produces qualitative differences in the sensitivity of a cable to geomagnetic storms. In particular, the same increase of the south-north component would produce an increase of the voltage or a decrease depending upon the direction of the current. A large increase of the voltage is the worst scenario, because the power supply may not be able to deliver the excess voltage required to feed the in-line components with the prescribed current, whereas it can easily reduce its voltage when necessary. Our finding that the voltage is proportional to the deviations of the magnetic field rather than to their time derivative implies that during a positive or negative magnetic field peak the voltage does not show both positive and negative voltage variations. Our analysis gives indications on the choice of the positive and negative side of the line to reduce the sensitivity of the systems to geomagnetic storms.  Indeed, in the strongest geomagnetic storm recorded to date, the Carrington event in 1859 \cite{Cliver:13}, the magnetic field underwent a sharp decrease, and this is the case also of the event that occurred on the 13th of March 1989 reported in Fig. \ref{Fig8}. In addition, there is experimental evidence that the strongest geomagnetic storms will be associated to a weakening of the geomagnetic field \cite{Mohanty:16}. Therefore, it would be worth designing cables where a decrease of the magnetic field is associated to a decrease of the required voltage, which does not harm system operations. In east-west cables, this would correspond to having the positive end westbound and the negative (or zero) eastbound, with the current in the cable going west-east. Note that if the direction of the current in the TAT-8 cable had been east-west instead of west-east, powering the cable during the magnetic storm of March 13 1989 would have required a voltage of 200 V in excess of the nominal value.

%a voltage of 200 V in excess of the nominal value would have been necessary to power the cable during the magnetic storm of March 13, 1989, and the required voltage may have been outside system specifications. 

\section{Conclusions}

We have analyzed the response of the voltage of the power supply of a transoceanic transmission system to geomagnetic field transients. We have shown that the voltage perturbation are proportional to the magnitude of the geomagnetic field perturbation and not to its time derivative. This result lead us to suggest design criteria to reduce the sensitivity of fiber optics cables to strong geomagnetic storm events.

% \bibitem{Zhang:14}
% Y.~Zhang, S.~Qiao, L.~Sun, Q.~W. Shi, W.~Huang, L.~Li, and Z.~Yang,
%   \enquote{Photoinduced active terahertz metamaterials with nanostructured
%   vanadium dioxide film deposited by sol-gel method,}
%   {\protect\JournalTitle{Optics Express}} \textbf{22}, 11070--11078 (2014).

% \bibitem{OSA}
% {Optical Society}, \enquote{{OSA Publishing},}
%   \url{http://www.osapublishing.org}.

% \bibitem{FORSTER2007}
% P.~Forster, V.~Ramaswamy, P.~Artaxo, T.~Bernsten, R.~Betts, D.~Fahey,
%   J.~Haywood, J.~Lean, D.~Lowe, G.~Myhre, J.~Nganga, R.~Prinn, G.~Raga,
%   M.~Schulz, and R.~V. Dorland, \enquote{Changes in atmospheric consituents and
%   in radiative forcing,} in \enquote{Climate Change 2007: The Physical Science
%   Basis. Contribution of Working Group 1 to the Fourth assesment report of
%   Intergovernmental Panel on Climate Change,}  S.~Solomon, D.~Qin, M.~Manning,
%   Z.~Chen, M.~Marquis, K.~B. Averyt, M.~Tignor, and H.~L. Miler, eds.
%   (Cambridge University Press, 2007).

% \end{thebibliography}

\end{document}